\documentstyle[12pt,epsf]{article}
\begin{document}
\title{Phantom Cosmology with Non-minimally Coupled Real Scalar Field}
\author{\large Zu-Yao Sun$^{1,2,4}$   \large You-Gen Shen$^{1,2,3}$}
\date{}
\maketitle \vspace*{6mm}

\footnotetext[1]{Shanghai Astronomical Observatory, Chinese
Academy of Sciences, Shanghai 200030, China; e-mail:
ygshen@center.shao.ac.cn.} \footnotetext[2]{National Astronomical
Observatories, Beijing
 100012, China.}
\footnotetext[3]{Institute of Theoretical Physics, Chinese Academy
of Sciences, Beijing 100080, China.}
\footnotetext[4]{Graduate
School of Chinese Academy of Sciences, Beijing 100039,  China.}

Please send proofs to: \hspace*{5mm}
\begin{minipage}[t]{10cm}
Prof. You-Gen Shen

Shanghai Astronomical Observatory,

80 Nandan Road,

Shanghai 200030, China.\\

e-mail: ygshen@center.shao.ac.cn

Tel: (86)(21)64386191-517

Fax: (86)(21)64384618
\end{minipage}
\newpage
\begin{abstract}We find that the expansion of the universe is accelerating by
analyzing the recent observation data of type $\textsc{I}a$
supernova(SN-Ia) .It indicates that the equation of state of the
dark energy might be smaller than -1,which leads to the
introduction of phantom models featured by its negative kinetic
energy to account for the regime of equation of state parameter
$w<-1$.In this paper the possibility of using a non-minimally
coupled real scalar field as phantom to realize the equation of
state parameter  $w<-1$ is discussed.The main equations which
govern the evolution of the universe are obtained.Then we rewrite
them with the observable quantities.
\end{abstract}
\section {INTRODUCTION}
\hspace*{7.5mm}Recent observations of the microwave background
show that the universe is almost flat[1,2]. And the astrophysical
data of type $\textsc{I}a$ supernova(SN-Ia) reveals that the
universe is currently undergoing a period of accelerating
expansion[3,4]. It follows immediately that there must be a kind
of dark energy, which makes up of two thirds of the energy density
and has negative pressure that can drive the accelerating
expansion of the universe. Many candidates for dark energy have
been proposed so far to fit the current observations. They are
cosmological constant, tachyon and a time varying scalar field
with positive or negative kinetic energy evolving in a specific
potential, referred to as ``quintessence'' or ``phantom'' and so
on. The major difference among these models are that they predict
different equation of state of the dark energy. As to cosmology
constant and ``quintessence'',the equation of state are confined
with in the range of $-1<w<-\frac{1}{3}$.Cosmologists have
proposed many models for quintessence cosmology,and discussed many
problems for it [5,6,7,8,9,10,11,12,13,14]. However, recent
observations do not exclude, but actually suggest a dark eneygy
equation of state $-1.38<w<-0.82$[15]. Therefore,some authors
investigated phantom field models that possess negative kinetic
energy and can realize $w<-1$ in their evolution.It has some
strange properties. For example,the energy density of phantom
energy increases with time.It also violates the dominant-energy
condition,which helps prohibit time machine and
wormholes.However,phantom is an interesting topic because it fits
current observations. A striking consequence of dark energy with
$w<-1$ is that our universe would
end in a ``Big Rip''[16,17].  \\
\hspace*{7.5mm}Several scalar-field models have been proposed for
phantom energy[16,18,19,\\20,21,22]. Theorists have also discussed
stringy phantom energy[23] and brane-world phantom energy[24]. In
this letter we take a non-minimally coupled real scalar field into
account,and show that it can be phantom.The information on phantom
may be determined with the observation data $r\left(z\right)$ from
the reconstruction equations.We also discussed the feasibility of
yielding the equation of state of phantom with the data
$r\left(z\right)$.  \\
\hspace*{7.5mm}Throughout the paper the units $G=c=1$ are used.\\

\section{FIELD EQUATION}

\hspace*{7.5mm}We start from the flat Robertson-Walker metric
\begin{equation}
ds^{2}=dt^{2}-a^{2}\left(t\right)\left(dr^{2}+r^{2}d\varphi_{1}^{2}
+r^{2}\sin^{2}\varphi_{1}d\varphi_{2}^{2}\right),
\end{equation}
where $a\left(t\right)$ is the scale factor of the universe.\\
\hspace*{7.5mm}The energy-momentum tensor of the scalar field is
given by
\begin{eqnarray}
\overline{T}_{\mu\nu}&=&\left(2\xi-1\right)\phi_{,\mu}\phi_{,\nu}
+\left(\frac{1}{2}-2\xi\right)g_{\mu\nu}\phi_{,\alpha}\phi^{,\alpha}
+2\xi\phi_{,\mu\nu}\phi-\frac{1}{2}\xi g_{\mu\nu}\phi\Box\phi
\nonumber\\&&+\xi
G_{\mu\nu}\phi^2+\frac{3}{2}\xi^2Rg_{\mu\nu}\phi^2+\left(1-3\xi\right)
g_{\mu\nu}V,
\end{eqnarray}
where $\Box$ is the d'Alembert operator, $G_{\mu\nu}$ is Einstein
tensor,$V$ is the potential of the field,
$R=6\left(a\ddot{a}+\dot{a}^2\right)/a^2$ is the Ricci scalar. For
the dust universe, the energy-momentum tensor of the
non-relativistic matter is
\begin{equation}
\widetilde{T}_{\mu\nu}=\rho_m U_{\mu}U_{\nu},
\end{equation}
where $\rho_m$ is the matter density and $U_{\mu}$ is the
4-velocity of the dust particles. Therefore, Einstein equations
can be written as
\begin{eqnarray}
H^2=\frac{\dot{a}^2}{a^2}&=&\frac{8\pi}{3}\left[\rho_m-\frac{1}{2}
\dot{\phi}^2+V+\frac{3}{2}\xi^2R\phi^2 \right.\nonumber\\&&\left.
+\xi\left(\frac{3}{2}\phi\ddot{\phi}-\frac{3}{2}
H\phi\dot{\phi}-3H^2\phi^2-3V\right)\right],
\end{eqnarray}
\begin{eqnarray}
\frac{\ddot{a}}{a}&=&-\frac{4\pi}{3}\left[\rho_m-2\dot{\phi}^2
-2V-3\xi^2R\phi^2\right.\nonumber\\&&\left.+
\xi\left(6\dot{\phi}^2+3\ddot{\phi}\phi+3H\phi\dot{\phi}
+6\frac{\ddot{a}}{a}\phi^2+6V\right)\right].
\end{eqnarray}
The coupled real scalar field contributes the energy density
$\rho_\Phi$ and pressure $p_\Phi$ as follows
\begin{eqnarray}
\rho_{\Phi}&=&-\frac{1}{2}
\dot{\phi}^2+V+\frac{3}{2}\xi^2R\phi^2\nonumber\\&&
+\xi\left(\frac{3}{2}\phi\ddot{\phi}-\frac{3}{2}
H\phi\dot{\phi}-3H^2\phi^2-3V\right),
\end{eqnarray}
\begin{eqnarray}
p_{\Phi}&=&-\frac{1}{2}
\dot{\phi}^2-V-\frac{3}{2}\xi^2R\phi^2\nonumber\\&&
+\xi\left(2\dot{\phi}^2+\frac{1}{2}\phi\ddot{\phi}+\frac{3}{2}
H\phi\dot{\phi}-\phi^2G_{1}^{1}+3V\right).
\end{eqnarray}
where $G_{1}^1$ is one component of Einstein tensor. We focus on
the strongly coupled case,then Eq.(4) and (5) can be simplified to
be
\begin{eqnarray}
H^2=\frac{\dot{a}^2}{a^2}=\frac{8\pi}{3}\left[\rho_m-\frac{1}{2}
\dot{\phi}^2+V+\frac{3}{2}\xi^2R\phi^2\right],
\end{eqnarray}
\begin{eqnarray}
\frac{\ddot{a}}{a}=-\frac{4\pi}{3}\left[\rho_m-2\dot{\phi}^2
-2V-3\xi^2R\phi^2\right].
\end{eqnarray}
In this case the equation-of-state for phantom field is
\begin{equation}
w=\frac{p_{\Phi}}{\rho_{\Phi}}=\frac{-\frac{1}{2}
\dot{\phi}^2-V-\frac{3}{2}\xi^2R\phi^2}{-\frac{1}{2}
\dot{\phi}^2+V+\frac{3}{2}\xi^2R\phi^2}
\end{equation}
It is clear that the strongly coupled field could realized the
equation-of-state $w<-1$ under the condition
\begin{equation}
V>\frac{1}{2}\dot{\phi}^2-\frac{3}{2}\xi^2R\phi^2
\end{equation}
 In Eq.(6) and
Eq.(7), the contributions from the field evolution $\dot\phi^2$
and coupled effect $\xi^2R\phi^2$ to the energy density and
pressure are proportional to $\left(d\phi/da\right)^2H^2a^2$ and
$R\phi^2\sim H^2\phi^2$, respectively. Provided that $\phi$ obeys
the power law $\phi=a^p$, the contribution of coupled effect and
field evolution are of the same order in terms of the universe
scale.Thus we can not neglect
either of them. \\

\section{EQUATION OF MOTION}
\hspace*{7.5mm} The Lagrangian density for the coupled real scalar
field $\Phi$ is
\begin{equation}
{\cal L}=\sqrt { - g} \left[ -\frac{{1}}{{2}}\partial _{\mu}  \Phi
\partial ^{\mu} \Phi + \frac{{1}}{{2}}\xi R\Phi^{2}
- V\left(\Phi\right)\right],
\end{equation}
where $\xi$ is a numerical factor. Now we decompose the field into
homogeneous parts and fluctuations as follows
\begin{equation}
\Phi=\phi\left(t\right)+\delta\phi\left(t,\vec{x}\right),
\end{equation}
Using Eq.(13) the Lagrangian density can be written as
\begin{eqnarray}
{\cal
L}&=&\sqrt{-g}\left[-\frac{1}{2}\left(\dot{\phi}+\delta\dot{\phi}\right)^{2}
+\frac{1}{2a^{2}}\left(\nabla \delta\phi
\right)^{2}\right.\nonumber\\&&\left.+\frac{1}{2}\xi
R\left(\phi+\delta\phi\right)^{2}
-V\left(\phi+\delta\phi\right)\right],
\end{eqnarray}
where the dot denotes the derivative with respect to t, $\nabla$
is the Laplace operator.\\
 \hspace*{7.5mm}The variation of the lagrangian density Eq.(14)
 yields the equation of motion of the field[25,26]
\begin{equation}
\ddot{\phi}+3H\dot{\phi}+\frac{d}{d\phi}\left[\frac{1}{2}\xi
R\phi^{2}-V\right]=0,
\end{equation}
for the homogeneous parts and
\begin{equation}
\delta\ddot{\phi}+3H\delta\dot{\phi}-\frac{\nabla^2}{a^2}\delta\phi+\xi
R\delta\phi-V^{''}\left(\phi\right)\delta\phi=0,
\end{equation}
for fluctuations.\\
\hspace*{7.5mm}The solution can be obtained in the following form
\begin{equation}
\delta\phi=\delta\phi_0
e^{\alpha\left(t\right)+i\vec{k}\cdot\vec{x}}.
\end{equation}
From the above, we can see that the fluctuations will grow
exponentially if $\alpha$ is real and positive. Thus the field is
unstable in such case. Substituting Eqs.(17) into Eq.(16) , we
obtain
\begin{equation}
\dot{\alpha}^2+3H\dot{\alpha} +\frac{k^2}{a^2}+\xi R-V^{''}=0.
\end{equation}
We assume that $\ddot{\alpha}\ll\dot{\alpha}^2$. In order for
$\alpha$ not to be real and positive, it should generally has the
relation
\begin{equation}
H^2-\frac{4}{9}\left(\frac{k^2}{a^2}+\xi R-V^{''}\right)<0,
\end{equation}
Then the coupled real scalar field becomes stable and may play the
role of phantom.
    Eq.(4),
Eq.(5), and Eq.(15) are the main equations which governing the
evolution of the universe.The term $\frac{1}{2}\xi R\phi^2$ in
Eq.(15), which is coming from the the ``coupled effect'' of the
field can be teated as an effective potential. It produces a
``centrifugal force'' and tends to drive $\phi$ away from zero.

\section{NUMERICALLY ANALYSIS}
\hspace*{7.5mm}Then we would numerically study the system in a
specific potential and obtain the results that might confirm our
qualitative analysis.\\
 \hspace*{7.5mm}To do so, we choose the
potential as
\begin{equation}
V\left(\phi\right)=V_0\left(1+\frac{\phi}{\phi_0}\right)exp\left(-\frac{\phi}{\phi_0}\right)
\end{equation}
Eq(8) could be rewritten as
\begin{equation}
H^2=H_i^2\left(\frac{\rho_{\Phi}}{\rho_{c,i}}+\Omega_{m,i}(\frac{a_i}{a})^3\right)
\end{equation}
 \hspace*{7.5mm}Where the subscript $i$ denotes the quantity at a initial time
$t_i$. $\rho_{c,i}$ is the critical density of the universe at
$t_i$,which is defined as $\rho_{c,i}=\frac{3H_i^2}{K}$,and
$K=8\pi$. $\Omega_{m,i}$ is the cosmic density for matter at
$t_i$.
 \\Introducing the new variables
\begin{equation}
X=\phi,\ \ \ \ Y=\frac{d\phi}{dt},\ \ \ \ N=ln\frac{a}{a_i},
\end{equation}
Then we can rewrite the main equations in the form
\begin{equation}
\frac{dX}{dN}=\frac{Y}{H_i}E(N)
\end{equation}
\begin{equation}
\frac{dY}{dN}=-3Y-\frac{E(N)}{H_i}\frac{d}{dX}(\frac{1}{2}\xi
RX^2-V)
\end{equation}
    Where
\begin{equation}
E(N)=(\frac{K\rho_\phi}{3H_i^2}+\Omega_{m,i}e^{-3N})^{-\frac{1}{2}}
\end{equation}
    We can get some insights into the evolution of the field by solving these equations.
We give a few of different quantities at the initial time,and take
different points as $t_i$,such as equipartition epoch and present
day.Then we find that the evolution of the field is not sensitive
to them.Thus,we can obtain the evolution of the equation of state
parameter $w$ roughly by taken some quantities not very strict.The
results are shown in Fig.1(non-minimally coupled case in which
$\xi$=50) and Fig.2(minimally coupled case in which $\xi$=0).They
are done by choosing $V_0$=100,$\phi_0$=1.

 \begin{center}
  \leavevmode
  \epsfbox{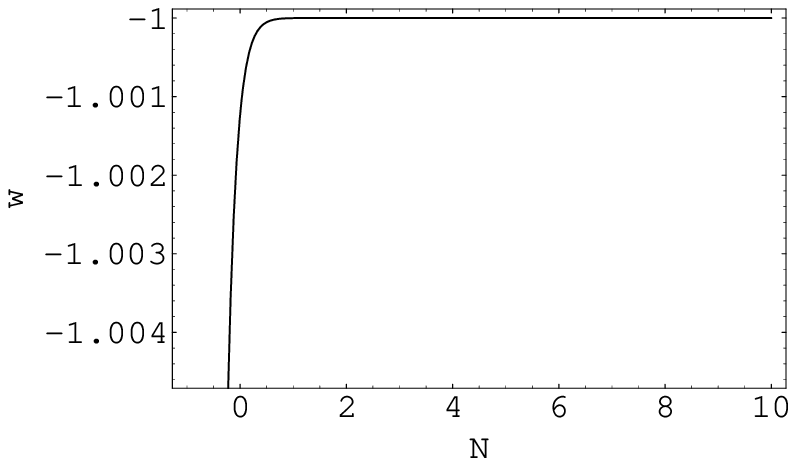}
\end{center}
Fig.1 The evolution of Phantom equation of state $w$ with
respect to N in non-minimally coupled case, for $\xi=50$.\\
\\
\\
 \begin{center}
  \leavevmode
  \epsfbox{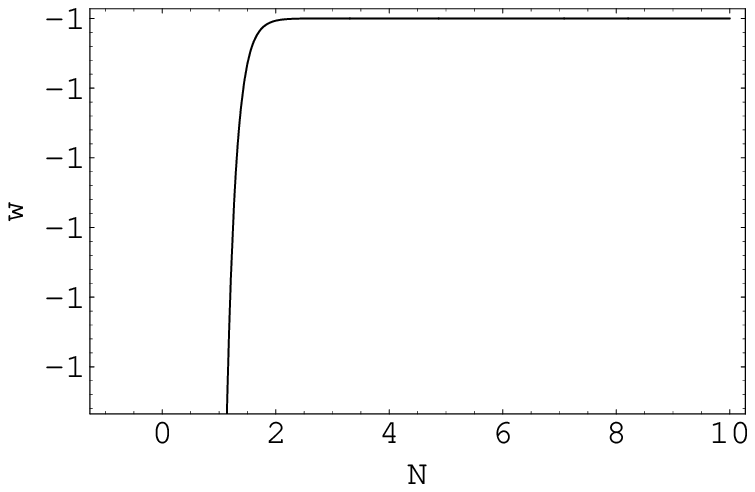}
\end{center}
Fig.2 The evolution of Phantom equation of state $w$ with respect
to N
in minimally coupled case, for $\xi=0$.\\
 \hspace*{7.5mm}From Fig.1 and Fig.2 we can see that the coupled
 constant $\xi$ could make the equation of state parameter $w$
 rise to -1 more quickly.

\section*{RECONSTRUCTION}
\hspace*{7.5mm}Now,we correlate the observable quantities with
non-observable quantities.\\
\hspace*{7.5mm}To do so,following the earlier study in this
field,we introduce the relations between $r\left(z\right), H_0$,
$\Omega_m$ and $a\left(t\right), H\left(t\right)$,
$\rho_m\left(t\right)$ as follows
 \begin{eqnarray}
1+z&=&\frac{1}{a},\ \ \ \ r\left(z\right)
=-\int^{t\left(z\right)}_{t_0}\frac{dt^{'}}{a\left(t^{'}\right)}
=\int^{z}_{0}\frac{dz^{'}}{H\left(z^{'}\right)},\nonumber\\&&\nonumber\\
H\left(z\right)&=&\frac{\dot{a}}{a}=\frac{1}{dr/dz},\ \ \ \
\rho_m=\Omega_m\rho_c=\frac{3\Omega_m}{8\pi}H_0^2\left(1+z\right)^3,
\\&&\nonumber\\\nonumber &&\frac{\ddot{a}}{a}=\frac{1}{\left(dr/dz\right)^2}
+\left(1+z\right)\frac{d^2r/dz^2}{\left(dr/dz\right)^3}.\nonumber
\end{eqnarray}
where $z$, $H_0$ and $\Omega_m$ are redshift, Hubble constant and
matter energy density, respectively. Using Eq.(26), we can obtain
the reconstruction equation
\begin{eqnarray}
 &&\dot{\phi}^2-\xi\left[2\phi\ddot{\phi}+2\dot{\phi}^2+2\phi^2\left(1+z\right)\frac{d^2r/dz^2}
{\left(dr/dz\right)^3}\right]\nonumber\\ &&=\frac{1}{4\pi}\cdot
\frac{\left(1+z\right)d^2r/dz^2}{\left(dr/dz\right)^3}+
\frac{3\Omega_m}{8\pi}H_0^2\left(1+z\right)^3,
\end{eqnarray}
\begin{eqnarray}
 && \frac{1}{2}\xi\left[\phi\ddot{\phi} -2\dot{\phi}^2-3H\phi
 \dot{\phi}-6H^2\phi^2-2\phi^2\left(1+z\right)\frac{d^2r/dz^2}
{\left(dr/dz\right)^3}-6V\right]\nonumber\\ &&+9\xi^2\phi^2
\left[\frac{2}{\left(dr/dz\right)^2}+\left(1+z\right)\frac{d^2r/dz^2}
{\left(dr/dz\right)^3}\right]+V\nonumber\\&&
=\frac{1}{8\pi}\left[\frac{3}{\left(dr/dz\right)^2}
+2\left(1+z\right)\frac{d^2r/dz^2}{\left(dr/dz\right)^3}\right]-
\frac{3\Omega_m}{16\pi}H_0^2\left(1+z\right)^3,
\end{eqnarray}
where $\dot{\phi}$ and $\ddot{\phi}$ are defined by
\begin{eqnarray}
\dot{\phi}&=&\frac{d\phi}{dt}=-\left(1+z\right)\frac{1}{dr/dz}\frac{d\phi}{dz},\nonumber\\&&\\
\ddot{\phi}&=&\frac{d^2\phi}{dt^2}=\left(1+z\right)^2\frac{1}{\left(dr/dz\right)^2}
\frac{d^2\phi}{dz^2}\nonumber \\&&
+\frac{d\phi}{dz}\left[\left(1+z\right)\frac{1}
{\left(dr/dz\right)^2}-\left(1+z\right)^2
\frac{1}{\left(dr/dz\right)^3}\frac{d^2r}{dz^2}\right].\nonumber
\end{eqnarray}
For strongly coupled case, Eq.[27] and Eq.[28] could be simplified
to
\begin{eqnarray}
 \dot{\phi}^2=\frac{1}{4\pi}\cdot
\frac{\left(1+z\right)d^2r/dz^2}{\left(dr/dz\right)^3}+
\frac{3\Omega_m}{8\pi}H_0^2\left(1+z\right)^3,
\end{eqnarray}
\begin{eqnarray}
 &&9\xi^2\phi^2
\left[\frac{2}{\left(dr/dz\right)^2}+\left(1+z\right)\frac{d^2r/dz^2}
{\left(dr/dz\right)^3}\right]+V\nonumber\\&&
=\frac{1}{8\pi}\left[\frac{3}{\left(dr/dz\right)^2}
+2\left(1+z\right)\frac{d^2r/dz^2}{\left(dr/dz\right)^3}\right]-
\frac{3\Omega_m}{16\pi}H_0^2\left(1+z\right)^3.
\end{eqnarray}
Therefore we can exam the model with the data
r$\left(z\right)$,and find the influence of coupled effect and
the potential of the field to the evolution of the universe.\\

\section*{CONCLUSION}
\hspace*{7.5mm}In conclusion, we have shown that it is possible
 to
use the coupled real scalar field as the phantom for the
accelerating of the universe, and it can realize the
equation-of-state $w<-1$ under the condition $
V>\frac{1}{2}\dot{\phi}^2-\frac{3}{2}\xi^2R\phi^2$. We also
compared the minimally coupled case and the non-minimally coupled
case by pictures of the variation of dark energy equation-of-state
$w$.Through the main equations which govern the evolution of the
universe,we can study the evolution of our universe. The coupled
term reveals the strong action between matter and dark energy. The
interaction between the two kinds of energy must be of great
importance in sometime of the universe evolution.
Therefore it should be seriously considered.\\
\hspace*{7.5mm}Eq.(6) and Eq.(7) show us the contributions to the
energy density and pressure that the coupled effect makes. It is
generally not negligible. Furthermore, provided that the field
evolution obeys the power law $\phi=a^p$,then the contributions of
the coupled effect and field evolution are of the same order in
terms of the universe scale.Current observations show that the
dark energy equation-of-state is in the range of $-1.38<w<-0.82$.
Thus, we must take phantom as a possible candidate of dark energy.
Because it can realize $w<-1$ while quintessence and cosmology
constant can realize $w\geq-1$.\\
\hspace*{7.5mm}The nature of the dark energy is still a
mystery.But we can tell the suitable one from
quintessence,cosmology constant and phantom by observations. We
may get the information of the phantom through the the observable
quantities by the reconstruction equations.The future data would
give us more proof to determine whether the dark energy is phantom
or not.
\\

\section*{ACKNOWLEDGEMENT}
\hspace*{7.5mm} The work has been supported by the National
Natural Science Foundation of China (Grant No. 10273017) and
Foundation of Shanghai Development for Science and
Technology(Grant No. 01JC14035).

\end{document}